\documentstyle[12pt]{article}

\catcode`\@=11
\long\def\@makefntext#1{ 
\protect\noindent \hbox to 3.2pt {\hskip-.9pt
$^{{\ninerm\@thefnmark}}$\hfil}#1\hfill} 

\def\thefootnote{\fnsymbol{footnote}}
 \def\@makefnmark{\hbox to 0pt{$^{\@thefnmark}$\hss}}  

\def\ps@myheadings{\let\@mkboth\@gobbletwo
\def\@oddhead{\hbox{} 
\rightmark\hfil\ninerm\thepage}
\def\@oddfoot{}\def\@evenhead{\ninerm\thepage\hfil 
\leftmark\hbox{}}\def\@evenfoot{}
\def\sectionmark##1{}\def\subsectionmark##1{}}

\begin{document}

\newcommand{\symbolfootnote}{\renewcommand{\thefootnote}
	{\fnsymbol{footnote}}}
\renewcommand{\thefootnote}{\fnsymbol{footnote}}
\newcommand{\alphfootnote}
	{\setcounter{footnote}{0}
	 \renewcommand{\thefootnote}{\sevenrm\alph{footnote}}}

\newcounter{sectionc}\newcounter{subsectionc}\newcounter{subsubsectionc}
\renewcommand{\section}[1] {\vspace{0.5cm}\addtocounter{sectionc}{1}
\setcounter{subsectionc}{0}\setcounter{subsubsectionc}{0}\noindent
{\bf\thesectionc. #1}\par\vspace{0.3cm}}
\renewcommand{\subsection}[1] {\vspace{0.5cm}\addtocounter{subsectionc}{1}
\setcounter{subsubsectionc}{0}\noindent {\it\thesectionc.\thesubsectionc.
#1}\par\vspace{0.3cm}}
\renewcommand{\subsubsection}[1]
{\vspace{0.5cm}\addtocounter{subsubsectionc}{1}
\noindent {\rm\thesectionc.\thesubsectionc.\thesubsubsectionc.
#1}\par\vspace{0.3cm}}
\newcommand{\nonumsection}[1] {\vspace{0.5cm}\noindent{\bf #1}
\par\vspace{0.3cm}}

\newcounter{appendixc}
\newcounter{subappendixc}[appendixc]
\newcounter{subsubappendixc}[subappendixc]
\renewcommand{\thesubappendixc}{\Alph{appendixc}.\arabic{subappendixc}}
\renewcommand{\thesubsubappendixc}
	{\Alph{appendixc}.\arabic{subappendixc}.\arabic{subsubappendixc}}

\renewcommand{\appendix}[1] {\vspace{0.5cm}
        \refstepcounter{appendixc}
        \setcounter{figure}{0}
        \setcounter{table}{0}
        \setcounter{equation}{0}
        \renewcommand{\thefigure}{\Alph{appendixc}.\arabic{figure}}
        \renewcommand{\thetable}{\Alph{appendixc}.\arabic{table}}
        \renewcommand{\theappendixc}{\Alph{appendixc}}
        \renewcommand{\theequation}{\Alph{appendixc}.\arabic{equation}}
        \noindent{\bf Appendix \theappendixc #1}\par\vspace{0.3cm}}
\newcommand{\subappendix}[1] {\vspace{0.5cm}
        \refstepcounter{subappendixc}
        \noindent{\bf Appendix \thesubappendixc. #1}\par\vspace{0.3cm}}
\newcommand{\subsubappendix}[1] {\vspace{0.5cm}
        \refstepcounter{subsubappendixc}
        \noindent{\it Appendix \thesubsubappendixc. #1}
	\par\vspace{0.3cm}}

\def\abstracts#1{{
	\centering{\begin{minipage}{30pc}\tenrm\baselineskip=12pt\noindent
	\centerline{\tenrm ABSTRACT}\vspace{0.3cm}
	\parindent=0pt #1
	\end{minipage} }\par}}

\newcommand{\bibit}{\it}
\newcommand{\bibbf}{\bf}
\renewenvironment{thebibliography}[1]
	{\begin{list}{\arabic{enumi}.}
	{\usecounter{enumi}\setlength{\parsep}{0pt}
\setlength{\leftmargin 1.25cm}{\rightmargin 0pt}
	 \setlength{\itemsep}{0pt} \settowidth
	{\labelwidth}{#1.}\sloppy}}{\end{list}}

\topsep=0in\parsep=0in\itemsep=0in
\parindent=1.5pc

\newcounter{itemlistc}
\newcounter{romanlistc}
\newcounter{alphlistc}
\newcounter{arabiclistc}
\newenvironment{itemlist}
    	{\setcounter{itemlistc}{0}
	 \begin{list}{$\bullet$}
	{\usecounter{itemlistc}
	 \setlength{\parsep}{0pt}
	 \setlength{\itemsep}{0pt}}}{\end{list}}

\newenvironment{romanlist}
	{\setcounter{romanlistc}{0}
	 \begin{list}{$($\roman{romanlistc}$)$}
	{\usecounter{romanlistc}
	 \setlength{\parsep}{0pt}
	 \setlength{\itemsep}{0pt}}}{\end{list}}

\newenvironment{alphlist}
	{\setcounter{alphlistc}{0}
	 \begin{list}{$($\alph{alphlistc}$)$}
	{\usecounter{alphlistc}
	 \setlength{\parsep}{0pt}
	 \setlength{\itemsep}{0pt}}}{\end{list}}

\newenvironment{arabiclist}
	{\setcounter{arabiclistc}{0}
	 \begin{list}{\arabic{arabiclistc}}
	{\usecounter{arabiclistc}
	 \setlength{\parsep}{0pt}
	 \setlength{\itemsep}{0pt}}}{\end{list}}

\newcommand{\fcaption}[1]{
        \refstepcounter{figure}
        \setbox\@tempboxa = \hbox{\tenrm Fig.~\thefigure. #1}
        \ifdim \wd\@tempboxa > 6in
           {\begin{center}
        \parbox{6in}{\tenrm\baselineskip=12pt Fig.~\thefigure. #1 }
            \end{center}}
        \else
             {\begin{center}
             {\tenrm Fig.~\thefigure. #1}
              \end{center}}
        \fi}

\newcommand{\tcaption}[1]{
        \refstepcounter{table}
        \setbox\@tempboxa = \hbox{\tenrm Table~\thetable. #1}
        \ifdim \wd\@tempboxa > 6in
           {\begin{center}
        \parbox{6in}{\tenrm\baselineskip=12pt Table~\thetable. #1 }
            \end{center}}
        \else
             {\begin{center}
             {\tenrm Table~\thetable. #1}
              \end{center}}
        \fi}

\def\@citex[#1]#2{\if@filesw\immediate\write\@auxout
	{\string\citation{#2}}\fi
\def\@citea{}\@cite{\@for\@citeb:=#2\do
	{\@citea\def\@citea{,}\@ifundefined
	{b@\@citeb}{{\bf ?}\@warning
	{Citation `\@citeb' on page \thepage \space undefined}}
	{\csname b@\@citeb\endcsname}}}{#1}}

\newif\if@cghi
\def\cite{\@cghitrue\@ifnextchar [{\@tempswatrue
	\@citex}{\@tempswafalse\@citex[]}}
\def\citelow{\@cghifalse\@ifnextchar [{\@tempswatrue
	\@citex}{\@tempswafalse\@citex[]}}
\def\@cite#1#2{{$\null^{#1}$\if@tempswa\typeout
	{IJCGA warning: optional citation argument
	ignored: `#2'} \fi}}
\newcommand{\citeup}{\cite}

\def\fnm#1{$^{\mbox{\scriptsize #1}}$}
\def\fnt#1#2{\footnotetext{\kern-.3em
	{$^{\mbox{\sevenrm #1}}$}{#2}}}

\font\twelvebf=cmbx10 scaled\magstep 1
\font\twelverm=cmr10 scaled\magstep 1
\font\twelveit=cmti10 scaled\magstep 1
\font\elevenbfit=cmbxti10 scaled\magstephalf
\font\elevenbf=cmbx10 scaled\magstephalf
\font\elevenrm=cmr10 scaled\magstephalf
\font\elevenit=cmti10 scaled\magstephalf
\font\bfit=cmbxti10
\font\tenbf=cmbx10
\font\tenrm=cmr10
\font\tenit=cmti10
\font\ninebf=cmbx9
\font\ninerm=cmr9
\font\nineit=cmti9
\font\eightbf=cmbx8
\font\eightrm=cmr8
\font\eightit=cmti8
%
%
\centerline{\tenbf VERIFICATION OF A NEW NON-LINEAR IV-EXPONENT:}
\baselineskip=16pt
\centerline{\tenbf SIMULATION OF THE 2D COULOMB GAS WITH LANGEVIN DYNAMICS.}
\vspace{0.8cm}
\centerline{\tenrm KENNETH HOLMLUND$^*$ and PETTER MINNHAGEN\footnote{\tenit
email: Kenneth.Holmlund@Physics.UmU.SE, Petter.Minnhagen@Physics.UmU.SE}}
\baselineskip=13pt
\centerline{\tenit Dept. of Theor. Physics, Ume{\aa} University,}
\baselineskip=12pt
\centerline{\tenit Ume{\aa}, S-90187, Sweden}
\vspace{0.9cm}
\abstracts{ It has recently been suggested\cite{expa} from scaling
arguments that the non-linear $IV$-exponent $a$, $V\propto I^a$, for
a two-dimensional superconductor is different from the exponent originally
suggested by Ambegaokar et al.(AHNS)\cite{AHNS}. The relation between the
new and the old exponent is $a=2a_{AHNS}-3$. The new scaling behaviour is
linked to the logarithmic vortex interaction and the long range time
tail which this gives rise to. Consequently one may expect that
the scaling behavior is generic for models which have these basic features.
The simplest model of this type is the two-dimensional Coulomb gas model with
Langevin dynamics. We here explicitly verify, through computer simulations,
that the $IV$-characteristics of this model indeed scales
according to the new scaling exponent $a$.
}
\vspace{0.8cm}

\rm\baselineskip=14pt
\vspace*{-0.3cm}  
The transition from the resistive state to the superconducting
state is for two dimensional (2D) superconductors of the
Kosterlitz-Thouless (KT) type.\cite{KT}
In the superconducting state the current-voltage ($IV$) characteristics
is non-linear, $V\sim I^a$, where the exponent $a$
is larger than one. An expression for $a$ was obtained by
Ambegaokar et al (AHNS) in ref.\cite{AHNS}. However, it has recently been
suggested that the exponent $a$ is linked to a critical dynamics of
the vortex fluctuations in the low temperature phase and that, as a
consequence,
it is different from the AHNS-value.\cite{expa} The relation between
the new and the old value of the exponent is $a=2a_{AHNS}-3$.\cite{expa}
The correctness of the new exponent was corroborated by simulations of the 2D
RSJ-model.\cite{expa}
Since the new exponent is linked to a critical
 dynamics and a critical scaling argument,
it is interesting to know
what ingredients are necessary to get this critical dynamics.
In the present paper we find a candidate for the
simplest generic model that has the required critical property\cite{holmlund}.

It is well-known that the KT-transition is due to unbinding of thermally
created
vortex-antivortex pairs and that the voltage is due to
flux flow of free vortices.\cite{minnhagen_rev} The interaction between
the vortices is logarithmic with distance. Our
hypothesis is that the critical dynamics is linked to the logarithmic
vortex interaction. The Coulomb gas interaction is logarithmic in
two dimensions and consequently the thermally created vortices and
antivortices can be described as a two dimensional gas of Coulomb
charges.\cite{minnhagen_rev}
The vortices are quantized and can have vorticity $s=\pm 1$, which
corresponds to positive
and negative Coulomb gas charges.
The total vorticity is always zero which means that the
Coulomb gas always has equally many positive and negative charges.
The vortices of a 2D superconductor obey a particular dynamics.\cite{AHNS}
Consequently the vortex part of a 2D superconductor can be modelled by a
Coulomb gas
with a particular dynamics for the Coulomb gas charges. However, if the
critical dynamics
is strongly linked to the logarithmic interaction, one may suspect that it is
relative
insensitive to the dynamical equation of the moving particles. This suggests
that
a 2D Coulomb gas with a simple dynamics would fall in the class of models which
have
the required critical property. In the present paper we show, through computer
simulations, that a 2D Coulomb gas with conventional Langevin dynamics indeed
belongs to
this class of models.

The equation of motion for the 2D Coulomb gas with Langevin dynamics is given
by
\begin{equation}
\frac{d {\bf r}(t)}{dt}=\frac{D}{T}{\bf F}_{tot} (t) +{\bf \eta}(t)
\end{equation}
where ${\bf r}$ is the position and ${\bf F}_{tot}$ is the total force acting
on it due
to all the other particles as well as any externally imposed force,
$D$ is the diffusion constant, $T$ is the temperature
(unit system such that the Boltzmann constant $k_B=1$), and
${\bf \eta}$ is a random force obeying
\begin{equation}
\langle \eta^\alpha (t) \eta^\beta (t')\rangle=2D\delta_{\alpha\beta}\delta
(t-t')
\end{equation}
where $\alpha$ and $\beta$ denote the Cartesian components. The charges are
circular
disks with diameter $r_0$ such that the force acting between two particles $i$
and $j$
with charges $s_i$ and $s_j$ respectively (units such that the charge is $s=\pm
1$) and
separated by the distance $r$ is given by
\begin{equation}
F_{ij}=s_is_j(\frac{1}{r}-\frac{1}{r_0}K_1(r/r_0))
\end{equation}
where $K_1$ is a modified Bessel function of order 1. Note that the force
between two particles
vanishes for $r=0$.
This means that the charge distribution of a particle is soft, which is in
accordance
with the precise vortex-Coulomb gas particle analogy.\cite{nylen}
In our simulations we choose a 2D box of length $L$ with periodic boundary
conditions
and a constant particle density $n$.
The results presented here are basically for $n=5\times 10^{-3}r_0^{-2}$ and
$L/r_0=226$ (
i.e. the total number of particles is $N=256$).
The simulations are done by discretizing  time with a time spacing
$\Delta t$ and introducing a Gaussian random noise ${\bf \eta}(t)$
acting independently
on each particle at each time step.
The equation of motion is then solved on the computer by
using a standard Euler integration method.\cite{holmlund}
The largest time sequences used in the present simulations are given
by $t/\Delta t \approx 5\times 10^6$. The $\Delta t$ chosen has
to be small enough to
ensure that the equation of motion is correctly solved yet as
large as possible in order to achieve as large time
sequencies as possible.\cite{holmlund}

We are calculating the average charge current $I_p$
as a function of an external force ${\bf F}_{ext}$.
The average charge current is given by
\begin{equation}
I_p=\langle \sum^{N}_{i=1}s_i \frac{dr_i (t)}{dt}\rangle=
\frac{D}{T}\langle \sum^{N}_{i=1} {\bf F}^{(i)}_{tot} (t) \rangle
\end{equation}
where the sum is over all particles $N$ and ${\bf F}^{(i)}_{tot}$
is the total force acting on the particle $i$ and the
brackets $\langle \rangle$ denote a time average.
The voltage $V$ in a 2D superconductor is due to the flux flow which means that
$V$ is proportional to $I_p$. The force acting on a vortex
is given by the Lorentz force and is consequently proportional to the current.
This means that the externally imposed current $I$ is proportional
to the external force $F_{ex}$. Thus the $IV$-characteristics for a
2D superconductor corresponds to the $F_{ex}I_p$-characteristics of
the Coulomb gas.
\begin{figure}[ht]
\setlength{\unitlength}{0.1bp}
\begin{picture}(3600,1943)(0,0)
\put(2008,51){\makebox(0,0){External Force, $F_{ex}$}}
\put(100,1071){%
\makebox(0,0)[b]{\shortstack{Charge Current, $I_{p}$}}%
}
\put(3417,151){\makebox(0,0){0.4}}
\put(3009,151){\makebox(0,0){0.3}}
\put(2433,151){\makebox(0,0){0.2}}
\put(1449,151){\makebox(0,0){0.1}}
\put(1132,151){\makebox(0,0){0.08}}
\put(724,151){\makebox(0,0){0.06}}
\put(540,1618){\makebox(0,0)[r]{0.1}}
\put(540,1162){\makebox(0,0)[r]{0.01}}
\put(540,707){\makebox(0,0)[r]{0.001}}
\put(540,251){\makebox(0,0)[r]{0.0001}}
\end{picture}
\fcaption{log-log plot of the $F_{ex}I_p$ characteristics ($T$ = 0.12, 0.14,
0.16, 0.18, 0.23, 0.26, from
bottom to top, respectively). The pluses represent the data and the dashed
lines are fits to the
form $AK_0(BF_{ex}) +C$ as described in text.}
\end{figure}
Figure 1 shows the obtained $F_{ex}I_p$-characteristics for a sequence of
temperatures
plotted as $\ln I_p$ against $\ln F_{ex}$.
As seen in the figure the data fall on straight lines
in the limit of small $F_{ex}$ which implies that $I_p \propto F^a_{ex}$ where
the slope of
the line gives the corresponding exponent $a$. In this way the non-linear
$F_{ex}I_p$-
exponent $a$ is obtained. As also seen in figure 1 the $F_{ex}I_p$-data can be
very well
fitted to the functional form $AK_0(BF_{ex}) +C\ln F_{ex}$ over an appreciable
range of $F_{ex}$,
where $A=1-a$, and $B$ and $C$ are constants.
This implies that $I_p = CF_{ex} \exp (AK_0(BF_{ex}))$ and this observation
somewhat
simplifies the determination of the exponent $a$.\cite{holmlund} The obtained
exponents
$a$, corresponding to the data in figure 1,
 are shown in figure 2 together with estimated error bars.
\begin{figure}[ht]
\begin{picture}(0,200)(0,0)
\end{picture}
\fcaption{The non-linear IV-exponent $a$ as a function of temperature (dashed
lines are guides to the eye).}
\end{figure}
According to the critical scaling theory these exponents $a$ should be related
to the static
charge density correlations.\cite{expa} The precise relation is
\begin{equation}
a=\frac{1}{T\tilde{\epsilon}}-1
\end{equation}
where $\tilde{\epsilon}$ is related to the Fourier transform
of the dielectric function $\hat{\epsilon}({\bf k})$. The dielectric function
is in turn
related to the charge density correlations by
\begin{equation}
\frac{1}{\hat{\epsilon} ({\bf k})}=1-\frac{2\pi L^2}{Tk^2}\langle \Delta
\hat{n} ({\bf k})
\Delta \hat{n} (-{\bf k}) \rangle
\end{equation}
where
\begin{equation}
\Delta n ({\bf r})=\sum^N_{i=1}s_i\delta ({\bf r}-{\bf r_i})
\end{equation}
is the charge density.
The Fourier transform of the dielectric function has the leading small $k$
dependence\cite{minnhagen_rev}
\begin{equation}
\frac{1}{\hat{\epsilon}({\bf
k})}=\frac{1}{\tilde{\epsilon}}\frac{k^2}{k^2+\lambda^{-2}}
\label{tde}
\end{equation}
where $\lambda$ is the screening length. This relation defines
$\tilde{\epsilon}$
in terms of $\hat{\epsilon} ({\bf k})$. The screening length $\lambda$ is
infinite
for temperatures below the KT transition temperature. This means that
$\tilde{\epsilon}=\hat{\epsilon} (k=0)$ in the low temperature phase.

In the simulations we calculate $\Delta \hat{n }({\bf k},t)$ for a long time
sequence for a given fixed value of ${\bf k}$. From such
a sequence we readily obtain the time average of the charge density
correlations $\langle \Delta \hat{n} ({\bf k}) \Delta \hat{n} (-{\bf k})
\rangle $ and from this
$1/\hat{\epsilon} ({\bf k})$. The value of $\tilde{\epsilon}$ is
then determined by using equation (\ref{tde}).
\begin{figure}[ht]
\begin{picture}(0,170)(0,0)
\put(200,20){\makebox(0,0){$\left|\bf{k}\right| \ \ \ \
\left[{2\pi\over{L}}\right]$}}
\put(18,110){%
\makebox(0,0)[b]{\shortstack{${1/{\hat{\epsilon}(\bf k)}}$}}
}
\end{picture}
\fcaption{The dielectric function as a function of $\left|\bf{k}\right|$ for
T=0.12,0.14,0.16,0.18,0.23,0.26, from
top to bottom, respectively, and the corresponding fits used to obtain
$\tilde{\epsilon}$.}
\label{ekfig}
\end{figure}
\begin{figure}[ht]
\begin{picture}(0,160)(0,0)
\put(200,20){\makebox(0,0){Coulomb Gas Temperature, T}}
\put(18,100){%
\makebox(0,0)[b]{\shortstack{${1/{\tilde{\epsilon}}}$}}
}
\end{picture}
\fcaption{$1/\tilde{\epsilon}$ obtained by fitting equation (\ref{tde}) to the
data in figure \ref{ekfig}.}
\end{figure}
This determination
of $\tilde{\epsilon}$ is shown in figure 3 and figure 4 gives the
values obtained . The scaling exponent $a=1/T\tilde{\epsilon}-1$
is now calculated from these values of $\tilde{\epsilon}$ and is compared to
the actual
exponent $a$ determined directly from our $I_pF_{ex}$-simulations in figure 2.
As seen in figure 2, the agreement with the scaling theory prediction is
excellent. According to the AHNS-theory\cite{AHNS} the exponent $a$ should
instead be given by
$a_{AHNS}=1/2T\tilde{\epsilon}+1=a/2+3/2$. This prediction is also shown in
figure 2.
As is apparent from figure 2, the AHNS-prediction does not agree with the
simulated exponents.
Thus figure 2 clearly verifies that the
new non-linear $IV$ exponent\cite{expa}
gives a correct description of the 2D Coulomb gas with
Langevin dynamics and that
the AHNS-prediction\cite{AHNS} is not correct for this model.
The same conclusion was obtained in case of the 2D RSJ-model.\cite{expa}
The present simulations suggests that the new scaling exponent is quite robust
and that
particles with a logarithmic interaction seems to be the most essential
feature for obtaining this new exponent.

According to the scaling theory\cite{expa}, the new critical exponent is linked
to a
long time tail in the charge density correlations. We have explicitly caculated
the long time behaviour of the charge density correlations function
\begin{equation}
\hat{g} ({\bf k},t)=\langle \hat{n} ({\bf k},t)\hat{n} (-{\bf k},0)\rangle
\end{equation}
{}From a practical point of view it is easier to converge the average for large
values of $t$
if one uses a $k$-value somewhat larger than the smallest possible $k=2\pi/L$.
Figure 5 shows
the result for a temperature ($T=0.18$) somewhat below the KT transition
($T_{KT}\approx 0.22$).
In this case a resonable convergence was achieved for the k-vector ${\bf
k}=\frac{2\pi}{L}(4,4)$.
The result is plotted as $t\hat{g}({\bf k},t)$ versus $t$. As seen in the
figure $t\hat{g}({\bf k},t)$
is constant to very good approximation for large $t$ which means that
$\hat{g}({\bf k},t)\propto 1/t$.
The result given in figure 5 shows that the time correlations for a small
finite $k$ fall to good approximation like $1/t$ over a very large
$t$-interval. This result
suggests that $\hat{g}({\bf k}=0,t)$ has a true $1/t$ tail for large $t$. \\
\begin{figure}[ht]
\begin{picture}(0,160)(0,0)
\put(200,21){\makebox(0,0){Time, t}}
\put(18,95){%
\makebox(0,0)[b]{\shortstack{$t \ \hat{g}({\bf k},t)$}}
}
\end{picture}
\fcaption{The charge density correlation function as a function of $t$. The
data clearly shows that
the charge density correlations decay as $1/t$ over a large time interval. Here
$t$ is in units of $\ \ r_0^2/D \ \ (\Delta t = 0.08 \ r_0^2/D)$.}
\end{figure}
In summary the results displayed in figures 1-4 demonstrate that the
2D Coulomb gas with Langevin dynamics has a non-linear $IV$-exponent $a$ which
is
distinctly different from the AHNS-prediction\cite{AHNS} and is instead given
by the scaling theory by Minnhagen at al.\cite{expa}. In addition the result
given in figure 5
is indeed consistent with a $1/t$-tail in the charge density correlations, as
required by
the scaling theory. We expect these findings to be quite general and so should
apply
to a 2D superconductor and a 2D Josephson coupled array.

{\em Acknowledgment}: The authors are indepted to Hans Weber for discussions
and for
making available to us his preliminary data on the lattice Coulomb gas which
clearly support
the results presented here. Support from the Swedish Natural Science Research
Council through
grant F-FU 0404-319 is greatfully  acknowledged.

\end{document}